\documentclass[11pt,twoside]{article}

\usepackage{asp2006}
\usepackage{epsf}
\usepackage{psfig}
\usepackage{lscape}

\begin{document}
\title{Feedback Processes: A Theoretical Perspective}   
\author{Mordecai-Mark Mac Low\altaffilmark{1}}   
\affil{Dept. of Astrophysics, American Museum of Natural History, 79th
Street at Central Park West, New York, NY 10024-5192, USA}    
\altaffiltext{1}{also Insitut f\"ur Theoretische Astrophysik, Zentrum
  f\"ur Astronomie der Universit\"at Heidelberg, and
  Max-Planck-Insitut f\"ur Astronomie}
\begin{abstract} 
I review the evidence for the importance of feedback from massive
stars at small and large scales. The feedback mechanisms include
accretion luminosity, ionizing radiation, collimated outflows, and
stellar winds. The good news is that feedback doesn't entirely prevent
the formation of massive stars, while the bad news is that we don't
know what does limit their masses. Feedback from massive stars also
influences their surroundings.  I argue that this does not produce a
triggering efficiency above unity, nor does it prevent lots of prompt
star formation in GMCs, though it may preserve massive remnants of the
clouds for many dynamical times.
\end{abstract}



\section{Small Scale Feedback and the Upper Mass Limit}

Massive stars influence their surroundings through a wide variety of
feedback mechanisms.  Their ionizing radiation
\citep*[e.g.,][]{vacca96} and line-driven winds \citep*{cak} have been
known for many years, while only more recently have collimated jets
and wide-angle outflows been observed from young, massive stars
\citep[e.g.,][]{bs05}.

The initial mass function for stars above a solar mass or so follows a
power law with slope -2.35 \citep{salpeter} up to an upper cutoff mass
of $m_{up} \sim 100$~M$_{\odot}$ \citep{f05}. The two most favored
ideas currently for what determines the slope of the IMF are turbulent
fragmentation of the parent cloud \citep{p07} or competitive
accretion \citep{bonnell}. What determines the value of the cutoff
mass $m_{up}$ remains unknown.  Limitation of accretion by
increasingly strong feedback in increasingly massive stars remains a
viable candidate mechanism \citep{zy07}.

In fact, the limitation of accretion by radiative feedback had long
been a severe problem in theories of massive star formation, because
in models of spherically symmetric accretion, radiation pressure on
standard interstellar dust prevents accretion on to stars with mass $M
> 30$~M$_{\odot}$ \citep{wolfire-cassinelli87}.  However, recent
models have called this conclusion into question because of better
treatment of dust properties, non-spherical accretion, and the high
accretion rates expected in high pressure cores \citep[e.g.,][]{mt03}
such as those that have been identified by the {\em Midcourse Space
Experiment} and {\em Spitzer Space Telescope} \citep{shirley03}.

\citet{ys02} performed two-dimensional, nested-grid, numerical models
of the collapse of slowly rotating, unmagnetized, molecular cores,
using a wavelength-dependent treatment of dust grain properties for a
mixture of small amorphous carbon grains, astrophysical silicates,
and, in colder regions, water and ammonia ice coatings.  They found
that a combination of decreased opacity, disk accretion, and focusing
of radiative losses through the polar cavity (the ``flashlight
effect'' first discussed by \citealt{yb99}) allowed formation of
35--45~M$_{\odot}$ stars.  \citet*{kmk05} used computations of
radiative transfer to demonstrate that the flashlight effect can be
further enhanced if a collimated gas outflow clears a cavity in the
envelope.

The efficacy of radiation in preventing accretion is further reduced
by two other effects.  First, \citet*{kkm05} showed that in three
dimensions, the radiation-driven bubbles that form when the radiation
pressure begins to repel gas are subject to Rayleigh-Taylor
instability.  Essentially, the radiation acts as a light fluid that is
trying to support the heavy gas.  As a result, the bubbles tend to
collapse, allowing additional accretion.  Second, in a magnetized
core, the photon bubble instability \citep{a92} can also act to allow
radiation to escape \citep*{tqy07}.  This occurs when compressive MHD
waves are amplified by radiation pressure, forming low-density regions
into which the radiation streams.

Although radiation does not prevent accretion entirely, it does heat
the surrounding core once a protostellar object has formed in the
center. The luminosity from accretion onto even a sub-solar mass
protostar can heat the surrounding envelope enough to prevent
fragmentation during the collapse of a centrally condensed core
\citep{k06}.  Adaptive mesh refinement simulations including radiative
transfer using a temperature dependent opacity show that fragmentation
is strongly suppressed when compared to models with an isothermal
equation of state \citep*{kkm07}.  At the time of publication, these
models had only reached 10~M$_{\odot}$, but more recent conference
reports have not yet given an upper mass limit.  Although this work
calls into question the competitive accretion scenario
\citep*{kmk05b}, it remains to be seen whether the two scenarios
converge to a common description of collapse when beginning from a
turbulent cloud rather than a centrally condensed core (see Clark or
Bonnell in this volume).

The good news here is that we seem to understand why feedback does not
entirely prevent massive star formation.  The bad news is that we
still don't know if it does or does not ultimately determine the upper
mass limit of around 100--150~M$_{\odot}$ \citep[e.g.,][]{f05}.
Several alternate explanations can be imagined.  For example, if
fragmentation cannot be prevented for cores above the upper mass limit
(perhaps because of insufficient central radiative heating), then the
upper mass limit would be determined by the largest mass reservoirs
available.  Alternatively, disk fragmentation could cut off accretion
around too massive stars.  Another suggestion is that massive stars
grow by direct collision, so that the stellar density is the limiting
factor. 

If feedback does act to cut off accretion, which mechanism is
dominant?  Jets and winds appear to evolve during the formation of
massive stars.  \citet{bs05} suggest an evolutionary sequence in which
the outflow type evolves as the central star accretes additional
mass. In this sequence, high mass protostellar objects with central
stars appearing as mid-to-early B stars have strongly collimated jets,
early~B to late~O stars that have already formed hypercompact H~{\sc
ii} regions start to show a line-driven wind in addition to the jet,
while early~O stars that have accreted most or all of their mass and
have ultracompact H~{\sc ii} regions (UCHRs) lose the jet and retain
only the wind.  This sequence could be symptomatic of disk destruction
and accretion cutoff in the final stages, a hypothesis supported by
the lack of observation of O~stars with collimated jets
\citep*{shepherd03,sollins+04,arce+07} or disks
\citep{cesaroni+07}. Whether the winds themselves can destroy the
disks remains an open question, however.

Jets have been suggested by \citet{nl07} as a way of supporting
cluster-forming cores.  However, the relatively low-resolution
numerical models they used do not yet resolve the question of whether
the turbulence driven in the cluster can actually cut off accretion.
Furthermore, their outflow-driven turbulence model predicts a
prominent break in the kinetic energy power spectrum at a scale close
to the outflow length.  Such a break has not yet been observed in a
molecular cloud. For example, \citet{om02} found no break in the
Polaris Flare---admittedly not a cluster forming core.  \citet{b07}
examine the velocity distribution produced by a single Mach 5 jet and
find very little supersonic material outside the head of the jet,
suggesting that jets may have difficulty driving the observed
supersonic turbulence in clouds.  On the other hand, \citet{matzner07}
offers an analytic treatment that argues that jets can indeed drive
the observed turbulence, consistent with \citet{nl07}.

Ionizing radiation from the growing central star offers another
mechanism for cutting off accretion.  \citet{keto02b} made the
important point that the formation of an H~{\sc ii} region around a
massive star does not cut off accretion until the Str\"omgren radius
grows to beyond the Bondi-Parker radius where the sound speed in the
ionized gas exceeds the local escape velocity.  \citet{keto07}
semi-analytically predicted the evolutionary sequence for stars
forming in rotating, collapsing, cores, using the similarity solution
derived by \citet{tsc84}.  A gravitationally confined,
quasi-spherical, hypercompact H~{\sc ii} region forms first, which
then transforms to a bipolar UCHR as accretion
increases the central star's mass and luminosity.  As the ionizing
luminosity increases further, only a remnant disk remains neutral, and
finally that evaporates as well.  The basic prediction of this model
that hypercompact H~{\sc ii} regions should show evidence of accreting
ionized gas has been borne out by observations both by \citet{keto02a,kw06}
and by others \citep{beltran06}.

If ionizing radiation is to cut off accretion, the final stage of
accretion will be through a photoevaporating disk.  This configuration
was first analytically described by \citet{h94} in the context of
solar mass star formation, while in the context of massive stars an
analytic treatment of the structure of the photoevaporating wind was
done by \citet*{lugo04}.  The best numerical model to date was done by
\citet{ry97} using a two-dimensional radiation gas dynamics code that
included an explicit treatment of dust scattering.  Dust scattering
leads to a factor of 3--4 increase in the photoevaporation mass loss
rate, which they show to be proportional to the stellar luminosity
$S_*^{0.58}$, close to the analytic prediction of \citet{h94}, though
flattening somewhat at photon luminosities $S_* > 10^{47}$~s$^{-1}$.
They also explicitly included line-driven stellar winds in some
models, demonstrating that they were collimated by the disk and
photoevaporating wind.

If UCHRs expand at the sound speed of ionized gas, $c_i \sim
10$~km~s$^{-1}$, they should have lifetimes of roughly $10^4$~yr. Less
than 1\% of an OB star's lifetime of a few megayears should therefore
be spent within an UCHR, so the same fraction of OB stars should now
lie within UCHRs.  However, \citet{wc89} surveyed UCHRs and found
numbers in our Galaxy consistent with over 10\% of OB stars being
surrounded by them, or equivalently, lifetimes $>10^5$~yr.

A number of explanations have been proposed for this lifetime problem,
including thermal pressure confinement in cloud cores, ram pressure
confinement by infall or bow shocks, champagne flows, disk
evaporation, and mass-loaded stellar winds (see \citealt{c99}).
Several of these explanations have basic problems that suggest they
likely cannot explain the lifetime problem. 

Confinement by thermal pressure of the surrounding molecular gas
requires pressures of $P/k \sim 10^8$--$10^9$~cm$^{-3}$~K
\citep*{drg95,gf96}. At typical molecular cloud temperatures of
10--100~K, this implies densities $n > 10^6$~cm$^{-3}$.  However, the
\citet{j1902} mass 
\begin{equation} \label{mjeans}
M_J = \left(\frac{4 \pi \rho}{3}\right)^{-1/2} \left(\frac{5 k T}{2 G
    \mu}\right)^{3/2} = 6.1 \mbox{ M}_{\odot}  \left(\frac{n}{10^6
    \mbox{ cm}^{-3}}\right)^{-1/2} \left(\frac{T}{100 \mbox{
      K}}\right)^{3/2}, 
\end{equation}
where we have assumed a mean mass per particle $\mu = 3.87 \times
10^{-24}$~cm$^{-3}$ appropriate for fully molecular gas with one
helium atom for every ten hydrogen nuclei.  Therefore cores massive
enough to form OB stars contain multiple Jeans masses and are thus
very likely to be freely collapsing (e.g.\ \citealt{mk04}).
The free-fall time
\begin{equation} \label{tff}
t_{\rm ff} = \left(\frac{3 \pi}{32 G \rho}\right)^{1/2} = (3.4 \times 10^4
\mbox{ yr}) \left(\frac{n}{10^6 \mbox{ cm}^{-3}}\right)^{-1/2}.
\end{equation}
Typical lifetimes of $>10^5$~yr would thus require massive cores to
last $>3 t_{\rm ff}$ at the hypothesized densities, rather than
dynamically collapsing.  Although these high pressures are indeed
observed, they are unlikely to occur in objects with lifetimes long
enough to solve the problem.  \citet{x96} instead proposed a variation
on this theme: confinement by turbulent rather than thermal
pressure. However, turbulent motions decay quickly, with a
characteristic timescale of less than a free-fall time under molecular
cloud conditions \citep*{sog98,m99}. Turbulent pressure would thus
have to be continuously replenished to maintain confinement for
multiple free-fall times, which would be difficult at such high
densities and small scales.

Another option is ram pressure confinement of UCHRs by infall of
surrounding gas.  However, this is unstable for two different physical
reasons. First, the density and photon flux will follow different
power laws in a gravitationally infalling region ionized from within,
so they can never balance each other in stable equilibrium
\citep{y86,h94}.  Either the ionized region will expand, or the infall
will smother the ionizing source. Second, the situation is Rayleigh-Taylor
unstable, as this option requires the rarefied, ionized gas of the
UCHR to support the infalling, dense gas.

Bow shock models \citep{v90,m91,ah06} require high values of ram pressure
$P_{ram} \propto n v_*^2$.  At $n=10^5$~cm$^{-3}$, a velocity of $\sim
10$~km~s$^{-1}$ is required \citep{v90}.  A star moving at such a high
velocity in a straight line would travel a parsec over the supposed
UCHR lifetime of $10^5$~yr, requiring a uniform-density region of mass
$>5 \times 10^3$~M$_{\odot}$ for confinement.  As collapse would occur
on the same timescale, more mass would actually be required to solve
the lifetime problem.

Expansion of an H~{\sc ii} region in a density gradient can drive
supersonic champagne flows down steep enough gradients
\citep{t79}. Two-dimensional models first studied expansion across
sharp density discontinuities \citep*{bty79}, but then examined other
configurations such as freely-collapsing cloud cores \citep*{ybt82},
clouds with power-law density gradients in spherical \citep*{ftb90},
and cylindrical \citep{gf96} configurations, and exponential density
gradients \citep{ah06}. Stellar wind combined with a champagne flow
down a continuous gradient was also modeled by \citet{ah06}.  This model
does fit well the observed velocity structure in some cometary UCHRs
such as G29.96$-0.02$. However, these models face the same timescale
problem as thermal pressure confinement models: regions dense enough
to explain the observations are gravitationally unstable, and collapse
on short timescales.

Another class of models relies on mass-loaded stellar winds to
reproduce the observed properties of UCHRs
\citep*{dwr95,rwd96,wdr96,l96}. In these models, an expanding stellar
wind entrains a distribution of small, self-gravitating,
pressure-confined clumps that take substantial time to evaporate.
These models can reproduce many of the basic features of the
observations including some line profiles \citep{dwr95}, core-halo,
shell \citep{rwd96}, and cometary and bipolar shapes \citep*{rwd98},
but at the cost of requiring an arbitrary distribution of pre-existing
clumps that cannot be self-consistently predicted.
 
Finally, \citet{m07} examines the expansion of H~{\sc ii} regions into
turbulent, self-gravitating gas.  At densities
high enough for massive stars to form, the expanding shell driven by
newly ionized gas quickly becomes gravitationally unstable
\citep{v88,mn93}, collapsing even more promptly than the surrounding
gas.  These regions of
secondary collapse in the shell may be externally ionized to form
objects with the properties of UCHRs.  As the shell expands to larger
sizes, new regions can form, extending the lifetime during which UCHRs
remain visible well beyond the expansion time of the original H~{\sc ii}
region.  Some combination of secondary collapse and disk evaporation
appears to me most likely to ultimately be the solution to the UCHR
lifetime problem.

\section{Large Scale Feedback}

The radiation, winds, and supernovae from massive stars also influence
their galactic environment.  Two questions of current interest are: 1)
Can compression driven by feedback trigger further star formation with
an efficiency above unity, allowing self-propagating star formation?
2) Can any of these forms of feedback support molecular clouds
for many dynamical times, reducing their star formation efficiency per
unit free-fall time, or, alternatively, giving a negative triggering
efficiency? 

Stellar feedback certainly produces compressive motions. The
suggestion that expanding H~{\sc ii} regions might compress
surrounding gas and trigger further massive star formation was first
made by \citet{el77} and \citet{ee78}.
They assumed that turbulent
velocities in the shell would be of the same order as the expansion
velocity.  \citet{oc81} and \citet{v83} assumed, on the other hand,
that the turbulent velocities in the shell would only be transonic.
This latter assumption was supported by analytic work by \citet{v88} and
numerical work by \citet{mn93}. 
The operation of this mechanism in three-dimensional simulations of
expansion into a turbulent medium is described by \citet{d05} using a
smoothed particle hydrodynamics code and by \citet{m07} using the
ZEUS-MP grid code \citep{n00}.  These models show that the clumpiness
of the turbulent medium allows the ionization front to blow out of the
collapsing core, while the denser regions resist ionization and
continue to collapse.  Neither model finds evidence for efficient
triggering---indeed in the more global model of \citet{d05} there is
some evidence for negative feedback---fewer stars forming in the
presence of the H~{\sc ii} region than otherwise.  \citet*{d07}
extended this result by examining an externally ionized collapsing
core, and found rather inefficient triggering in that case as well.
These low or negative triggering efficiencies occur when the
acceleration and stirring of the gas by a shock wave expanding into a
clumpy medium dominates over direct compression.

Supernovae compress gas on even larger scales than H~{\sc ii} regions,
so they are also often invoked as triggering agents.  \citet{jm06}
simulated supernova-driven turbulence using the Flash adaptive mesh
refinement code \citet{fryxell} on a grid covering 500~pc on a side in
the plane and 5~kpc up and down vertically, with 2~pc resolution
within 200~pc of the plane.  When the galactic fountain flow had
reached a steady state, the density, pressure, and velocity dispersion
of the gas in two-zone cubes was compared to the
turbulent Jeans criterion for collapse, to measure the expected star
formation rate in the flow.  If a generous local star formation
efficiency of 30\% is assumed for the unstable gas, they find that the
resulting star formation rate is a full order of magnitude below the
value required to generate the input supernova rate.

These results on triggering by ionizing radiation and supernova
explosions taken together suggest that triggering does indeed occur,
but that it is only a 10\% effect rather than the primary mechanism
controlling star formation (although it may be more important than
this for massive star formation in particular).  A recent
observational test supports this result. \citet{mizuno07} used the
NANTEN telescope to survey dense molecular gas interacting with 23
southern H~{\sc ii} regions within 4~kpc of the Sun.  The detected
clouds were divided into regions adjacent or not adjacent to the
H~{\sc ii} regions. The adjacent regions indeed contained more
protostellar objects, and ones with significantly higher far-infrared
luminosities luminous protostellar objects.  However, when the effect
is integrated over the galaxy, it appears to be only a 10--30\%
effect. 

If triggering is not that effective, is feedback good at the
converse---supporting molecular clouds against collapse?  The question
of how long-lived molecular clouds remains very controversial, with
answers ranging over an order of magnitude from one to as long as ten
free-fall times. Long time scales are argued for on the basis of
either magnetic \citep*{mtk06} or turbulent \citep*{kmm06} support,
while short time scales depend on sweeping by shock compression
\citep*{bhv99,hbb01,p07} or direct gravitational instability and collapse
\citep*{lmk05,e07}. 

\citet{m02} argued that turbulent support of molecular clouds can be
maintained by H~{\sc ii} region expansion within the clouds, for
clouds with a mass greater than $3.7 \times 10^5$~M$_{\odot}$.  His
analysis suggests that such large clouds reach a balance between
energy injection and decay in a turbulent cascade.  \citet*{kmm06} used
this analysis and made the additional assumption that giant molecular
clouds are spherical objects with homologous, power-law distributions
of density, energy injection, and other relevant quantities.  Using a
star formation law \citep{km05} derived from the turbulent collapse
models of \citet*{vbk03}, they find that large molecular clouds remain
quasi-stable for as long as 20--40~Myr.  Similar work was performed by
\citet{hs07} in the context of the Orion Nebula Cluster.

However, \citet{e07} reaches opposite conclusions working from almost
the same assumptions.  He argues that giant molecular clouds never
have spherical, homologous shapes, but rather are filamentary objects
containing quasi-spherical cluster-forming cores of order
$10^4$~M$_{\odot}$.  In his picture, these dense cores are dispersed
by H~{\sc ii} region expansion within 2--4~Myr, but magnetically
supported envelopes that are not participating in gravitational
collapse may indeed survive for much longer. 

The observational evidence has not yet clarified the picture either. A
study by Fukui's group of the associations between young clusters,
H~{\sc ii} regions, and molecular gas observed by NANTEN, reported in
\citet{blitz06}, suggests that GMCs remain without visible H~{\sc ii}
regions for about 7~Myr---though \citet{gruendl07} used {\em Spitzer}
to show that they do have young stellar objects.  About another 14~Myr
passes before visible clusters appear, and then 6~Myr remain before
the molecular gas is cleared away, for a total of as much as 27~Myr.
They also note that the average mass of the observed objects increases with
the stage of life.  They interpret this as suggesting continuing
accretion onto the clouds over their lifetimes.

On the other hand, \citet{t07} have used global imaging of nearby
galaxies from the {\em Spitzer} IR Nearby Galaxies Survey \citep{k03}
and The H~{\sc i} Nearby Galaxy Survey \citep{w07} to measure the
average delay between the peak of H~{\sc i} emission and the peak of
star formation as measured by 24~$\mu$m emission from warm dust in
regions of massive star formation. They find values of 2--4~Myr for
the delay, with no values above 6~Myr in their sample of over 20
galaxies. 

While the peak of 24~$\mu$m emission does not yield the total lifetime
of the molecular clouds, it does measure the period of strongest star
formation.  The results of \citet{t07} suggest that star formation
begins promptly and strongly within a few million years of the onset
of gravitational instability as traced by H~{\sc i}
emission. The timescales cited by \citet{blitz06} make no reference to
the strength of star formation during any particular period, though,
so perhaps the two observations are consistent with the theoretical
picture of cloud support by H~{\sc ii} region expansion: if clouds
collapse promptly, forming many young stars, the result could be well
supported clouds with low ongoing star formation efficiency.
Investigation of this scenario looks like a fruitful direction for
further research.

\acknowledgements I thank the organizers for their invitation to
speak, the Deutscher Akademischer Austausch Dienst and the
Max-Planck-Gesellschaft for stipends in support of my stay in
Heidelberg, and the National Science Foundation under grants
AST03-07793 and AST06-07111 for partial support of my research.


\end{document}